\documentstyle[emulateapj,psfig,natbib,amsfonts,amsmath,graphicx]{article}
\def\grb{GRB\,020127}
\def\ra#1#2#3{#1$^{\rm h}$#2$^{\rm m}$#3$^{\rm s}$}
\def\dec#1#2#3{$#1^\circ#2'#3''$}

\def\chandra{{\it Chandra X-ray Observatory}}

\def\hst{{\it Hubble Space Telescope}}
\def\hete{{\it High Energy Transient Explorer II}}

\def\ociw{1}
\def\prince{2}
\def\hubble{3}
\def\psu{4}
\def\cit{5}
\def\nrao{6}

\begin{document}
 
\title{\large The ERO Host Galaxy of GRB\,020127: Implications for the
Metallicity of GRB Progenitors}

\author{
E.~Berger\altaffilmark{\ociw,}\altaffilmark{\prince,}\altaffilmark{\hubble},
D.~B.~Fox\altaffilmark{\psu},
S.~R.~Kulkarni\altaffilmark{\cit},
D.~A.~Frail\altaffilmark{\nrao},
and S.~G.~Djorgovski\altaffilmark{\cit}
}

\altaffiltext{\ociw}{Observatories of the Carnegie Institution
of Washington, 813 Santa Barbara Street, Pasadena, CA 91101}

\altaffiltext{\prince}{Princeton University Observatory,
Peyton Hall, Ivy Lane, Princeton, NJ 08544}

\altaffiltext{\hubble}{Hubble Fellow}

\altaffiltext{\psu}{Department of Astronomy and Astrophysics,
Pennsylvania State University, 525 Davey Laboratory, University
Park, PA 16802}

\altaffiltext{\cit}{Division of Physics, Mathematics and Astronomy,
105-24, California Institute of Technology, Pasadena, CA 91125}

\altaffiltext{\nrao}{National Radio Astronomy Observatory, Socorro,
NM 87801}

\begin{abstract}
We present optical and near-IR observations of the host galaxy of
GRB\,020127, for which we measure $R\!-\!K_s=6.2$ mag.  This is only
the second GRB host to date, which is classified as an ERO.  The
spectral energy distribution is typical of a dusty starburst galaxy,
with a redshift, $z\approx 1.9$, a luminosity, $L\approx 5$ L$^*$, and
an inferred stellar mass of $M_*\sim 10^{11}-10^{12}$ M$_\odot$, two
orders of magnitude more massive than typical GRB hosts.  A comparison
to the $z\sim 2$ mass-metallicity ($M$-$Z$) relation indicates that
the host metallicity is about $0.5$ Z$_\odot$.  This result shows that
at least some GRBs occur in massive, metal-enriched galaxies, and that
the proposed low metallicity bias of GRB progenitors is not as severe
as previously claimed.  Instead we propose that the blue colors and
sub-$L^*$ luminosities of most GRB hosts reflect their young starburst
populations.  This, along with the locally increased fraction of
starbursts at lower stellar mass, may in fact be the underlying reason
for the claimed low metallicity bias of $z\lesssim 0.2$ GRB hosts.
Consequently, GRBs and their host galaxies may serve as a reliable
tracer of cosmic star formation, particularly at $z\gtrsim 1$ where
the $M$-$Z$ relation is systematically lower, and the star formation
rate is dominated by sub-$L^*$ galaxies.
\end{abstract}

\keywords{gamma-rays:bursts --- galaxies:starburst}

\section{Introduction}
\label{sec:intro}

Observations of the host galaxies of long-duration $\gamma$-ray bursts
(GRBs) have led to the general consensus that they are faint and blue.
This has been interpreted as evidence for intense star formation
activity, as well as low metallicity and stellar mass
\citep{fjm+03,fdm+03,chg04,fls+06,sgb+06}.  In particular, a
comparison of $z\lesssim 0.2$ GRB hosts to Sloan Digital Sky Survey
galaxies suggests that GRBs occur preferentially in low metallicity
galaxies with $Z\sim 0.1-0.5$ Z$_\odot$ \citep{sgb+06}.  These authors
also propose an upper metallicity cutoff of about 0.15 Z$_\odot$ for
typical GRBs (i.e., with $E\sim 10^{51}$ erg).  At higher redshifts,
$z\sim 0.5-1$, a comparison to the host galaxies of GOODS supernovae
indicates that GRBs tend to occur in lower luminosity galaxies (by
about 1 mag) and therefore presumably at lower metallicity
\citep{fls+06}.  Metallicities measured from afterglow absorption
spectra (at $z>2$) span a wider range, $Z\sim 0.05-0.5$ Z$_\odot$
(e.g., \citealt{bpc+06,pro06}).  Theoretical studies have also argued
for ``low'' metallicity of GRB progenitors \citep{mw99}, although we
stress that the definition is not well-quantified, and black hole
remnants can apparently be formed even above solar metallicity
\citep{hfw+03}.

These observations raise two crucial and related questions: (i) {\it
Is there in fact a low metallicity bias for GRB progenitors, in both
an absolute sense and relative to the average metallicity at any
redshift?}  (ii) {\it Can GRBs and their host
galaxies be used as a representative tracer of star formation at high
redshift ($z\gtrsim 1$)?}  The answer to these questions is of great
importance given the unique ability of GRBs to probe the interstellar
medium and star forming environments of galaxies over a wide redshift
range, extending to $z>6$ \citep{bcc+06}.  It is also crucial to
understand the effect of metallicity in light of the redshift
evolution of the average metallicity and mass function of galaxies
(e.g., \citealt{esp+06}).  Equally important, the metallicity range
for GRB progenitors impacts our understanding of the progenitor
population and GRB formation scenarios \citep{mw99,hfw+03}.

The prevalence of GRBs in blue galaxies may also reflect an
observational bias against dusty galaxies due to obscuration of the
optical afterglow.  Two lines of evidence suggest that this may not be
a significant problem.  First, the host galaxies of the ``optically
dark'' GRBs (those with clear evidence for dust obscuration in the
optical band) are typically not redder than the hosts of
optically-bright GRBs \citep{bck+03,fdm+03}.  A notable exception is
GRB\,030115, with a dust-reddened afterglow and a host galaxy with
$R\!-\!K\approx 5.4$ mag \citep{lfr+06}.  Second, radio,
submillimeter, and IR observations have not led to a preferential
detection of the dark GRB hosts \citep{bbt+03,bck+03,lcf+06}, despite
an expectation that these galaxies should have obscured star
formation.  Conversely, the host galaxies with long-wavelength
detections (and hence obscured star formation), have blue colors
typical of the GRB host population as a whole \citep{bck+03}.

Here we present optical and NIR observations of GRB\,020127, which
reveal that the host galaxy is an extremely red object (ERO) with
$R\!-\!K_s=6.2$ mag.  The GRB position is obtained from X-ray and
radio observations of the afterglow.  The host SED is best fit by an
obscured starburst galaxy template at $z\approx 1.9$, with a stellar
mass\footnotemark\footnotetext{We use Vega magnitudes (unless
otherwise noted), and the standard cosmology with $H_0=71$ km s$^{-1}$
Mpc$^{-1}$, $\Omega_m=0.27$, and $\Omega_\Lambda= 0.73$.} of $\sim
10^{11.5}$ M$_\odot$ and an inferred metallicity \citep{esp+06} of
about 0.5 Z$_\odot$, suggesting that GRB progenitors can in fact exist
at near-solar metallicity, and may occur in the most massive galaxies 
at $z>1$.

\section{Afterglow and Host Galaxy Observations}
\label{sec:obs}

\grb\ was discovered by the \hete\ satellite on 2002 Jan.~27.875 UT,
with a positional accuracy of $8'$ radius \citep{gcn1229}.  The burst
duration and fluence in the $2\!-\!400$ keV band are $18$ s and
$2.7\times 10^{-6}$ erg cm$^{-2}$ \citep{slk+05}.

Optical observations did not uncover an afterglow candidate, with
$R,I\!>\!19.5$ mag at 4.4 and 8.2 hr after the burst, respectively
\citep{gcn1230}, and $R\!>\!21.5$ mag at 3.1 hr over $75\%$ of the
error circle \citep{gcn1234}.  The Galactic extinction in the
direction of \grb\ is low, $E(B-V)=0.048$ mag \citep{sfd98}.

\subsection{X-ray, Radio, and Optical/NIR Imaging}

We initiated two 10 ks observations with the \chandra, $4.14$ and
$14.64$ d after the burst, to identify the fading X-ray afterglow.
The data were obtained with the Advanced CCD Imaging Spectrometer
(ACIS) and reduced and analyzed using the CIAO software
package\footnotemark\footnotetext{http://cxc.harvard.edu/ciao}.  A
comparison of the two epochs reveals three fading sources not
associated with bright stellar counterparts.  Of these, only one is
detected in both epochs, and has faded with high significance, with
2.6 and 1.0 count ks$^{-1}$ ($0.3-7$ keV), respectively.  Using a
Galactic column of $3\times 10^{20}$ cm$^{-2}$ \citep{sfd98} and a
photon index of $1.5\pm 0.4$, we find fluxes of $(5.5\pm 0.1) \times
10^{-14}$ and $(2.1\pm 0.7)\times 10^{-14}$ erg cm$^{-2}$ s$^{-1}$,
respectively, corresponding to a power law decay rate, $F_\nu\propto
t^{-0.8}$, typical of GRB afterglows.  The position of the X-ray
afterglow candidate is $\alpha$=\ra{08}{15}{01.42},
$\delta$=\dec{+36}{46}{33.9} (J2000), with an uncertainty of about
$1''$ in each coordinate.

We observed this source with the Very Large Array
(VLA\footnotemark\footnotetext{The VLA is operated by the National
Radio Astronomy Observatory, a facility of the National Science
Foundation operated under cooperative agreement by Associated
Universities, Inc.}) at a frequency of 8.46 GHz on 2002 Feb.~14.20,
16.23, 21.97, and Mar.~18.10 UT.  The data were processed using AIPS.
In the first observation we detect an object with a flux of $222\pm
63$ $\mu$Jy, coincident with the X-ray position, at
$\alpha$=\ra{08}{15}{01.42}, $\delta$=\dec{+36}{46}{33.45} (J2000),
$\pm 0.03''$ in each coordinate.  Subsequent observations reveal that
the object has faded, with $3\sigma$ limits of 150 $\mu$Jy (Feb.~16.23
and 21.97) and 65 $\mu$Jy (Mar.~18.10), suggesting that this is the
radio afterglow of \grb.

Optical observations were obtained with the Large Format Camera (LFC)
on the 200-inch telescope at the Mt.~Palomar Observatory on 2002
Feb.~4.29 and Feb.~6.34 UT for a total of 4300 s ($g'$), 2400 s
($r'$), and 1800 s ($i'$).  The images were bias-subtracted,
flat-fielded, and coadded using IRAF, and photometry was performed
relative to SDSS.  At the position of the X-ray and radio afterglows
we detect a faint object with $i'=23.89\pm 0.13$ mag, $r'>23.5$ mag,
and $g'>25.8$ mag; limits are $2\sigma$ and magnitudes are in the AB
system.  We identify this object as the host galaxy of \grb.  Further
observations were obtained with the Echellete Spectrograph and Imager
(ESI) on the Keck II 10-m telescope on 2002 Mar.~13.28 UT in $R$ (1500
s) and on 2003 Feb.~28.36 UT in $I$ (600 s).  The data were reduced as
described above.  We measure $I=23.56\pm 0.10$ mag and $R=24.73\pm
0.15$ mag.

\begin{figure}
\epsscale{1}
%\plotone{fig1.eps}
\centerline{\psfig{file=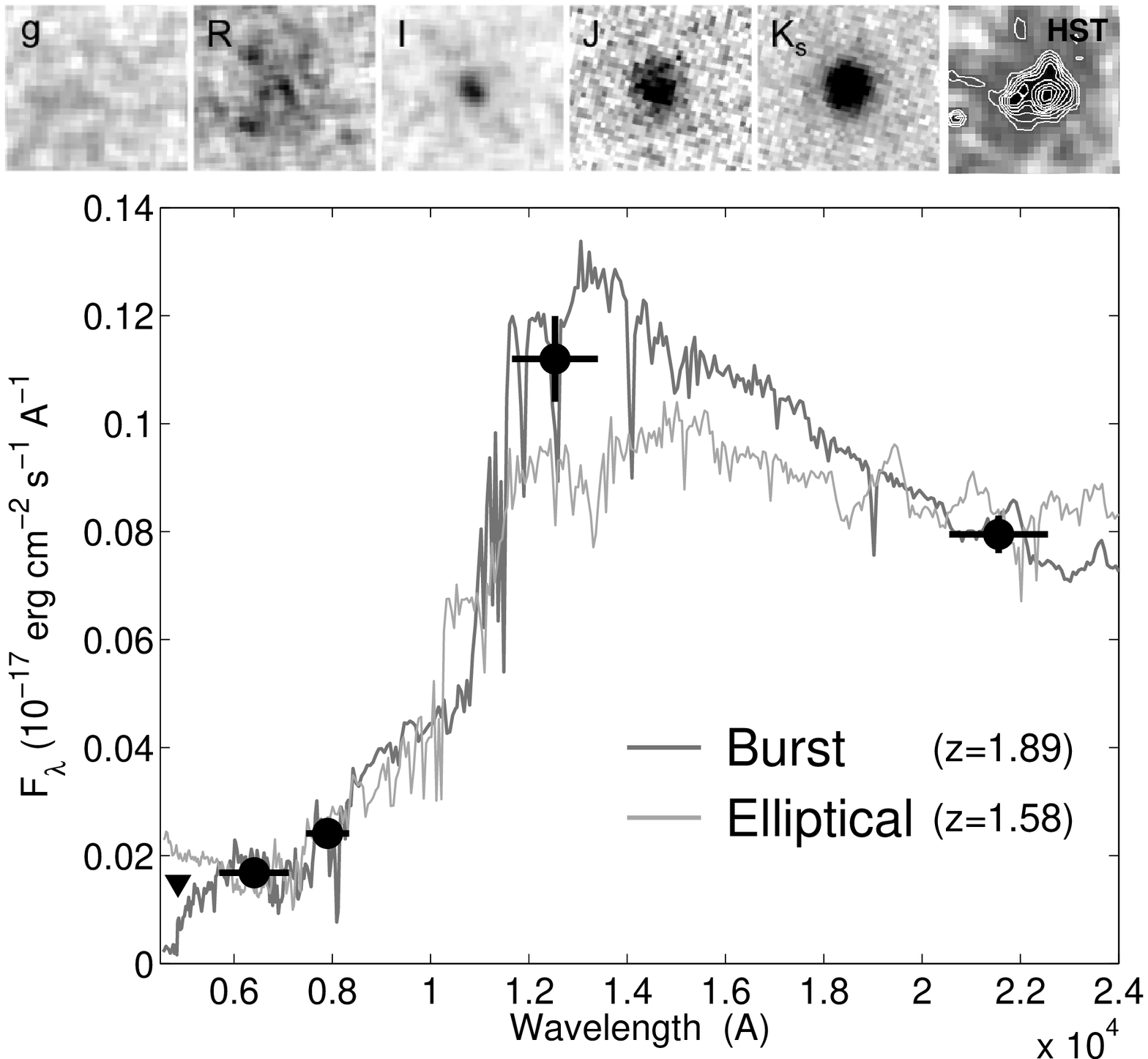,width=6.5in}}
\caption{Spectral energy distribution of the host galaxy of \grb\
along with the best fit starburst and elliptical galaxy models.  The
combination of a blue $J-K_s$ color and a deep upper limit in the
$g$-band indicates that the starburst template is a much better fit
compared to an elliptical galaxy.  The top panel shows the host galaxy
in each of the five observed bands.
\label{fig:sed}}
\end{figure}

NIR observations in $K_s$ and $J$ were obtained with the Near
Infra-Red Camera (NIRC) on the Keck I 10-m telescope on 2002
Oct.~13.63 and Nov.~17.48 UT, respectively, for a total of 1080 s in
each filter.  The individual frames were dark-subtracted,
flat-fielded, and corrected for bad pixels and cosmic rays using
custom IRAF routines.  Photometry was performed relative to the
standard star Feige 16.  The host galaxy has $J=20.37\pm 0.10$ mag and
$K_s=18.54\pm 0.05$ mag.  Optical/NIR images of the host are shown in
Figure~\ref{fig:sed}.

Finally, on 2002 Apr.~6 UT we observed the host galaxy with the \hst\
using the Space Telescope Imaging Spectrograph (STIS) as part of
program GO 9180 (PI: Kulkarni).  A total of 4868 s were obtained with
the CL filter.  We processed and combined the individual exposures
using the IRAF task {\tt drizzle} \citep{fh02}, with {\tt pixfrac=0.8}
and {\tt pixscale=0.5}.  At the position of the afterglow we detect an
extended object with an AB magnitude of $24.8\pm 0.1$ mag
(Figure~\ref{fig:sed}).

\subsection{Optical Spectroscopy} 

We obtained optical spectra of the host galaxy using ESI on four
separate occasions.  The data were reduced using custom IRAF routines
to bias-subtract, flat-field, and rectify the ten individual echelle
orders.  Sky subtraction was performed using the method and software
described in \citet{kel03}.  Wavelength calibration was performed
using CuAr and HgNeXe arc lamps and air-to-vacuum and heliocentric
corrections were applied.  The spectrum covers the range $0.39$ to
$1.05$ $\mu$m at a resolution of 11.5 km s$^{-1}$.  We detect
continuum emission beyond 0.6 $\mu$m, but no emission or absorption
lines are clearly identified.

A 5400 s spectrum of the host was obtained with the Low Resolution
Imaging Spectrometer (LRIS) on the Keck I 10-m telescope on 2004
Apr.~22 UT.  The wavelength coverage is $0.36$ to $0.96$ $\mu$m with a
resolution of $3.3$ \AA\ ($0.36\!-\!0.58$ $\mu$m) and $5.6$ \AA\
($0.58\!-\!0.96$ $\mu$m).  The data were reduced as described above.
As in the ESI spectra, we detect the host continuum but no emission or
absorption features.

\section{Host Galaxy Properties}
\label{sec:res}

\begin{figure}
\epsscale{1}
%\plotone{fig2.ps}
\centerline{\psfig{file=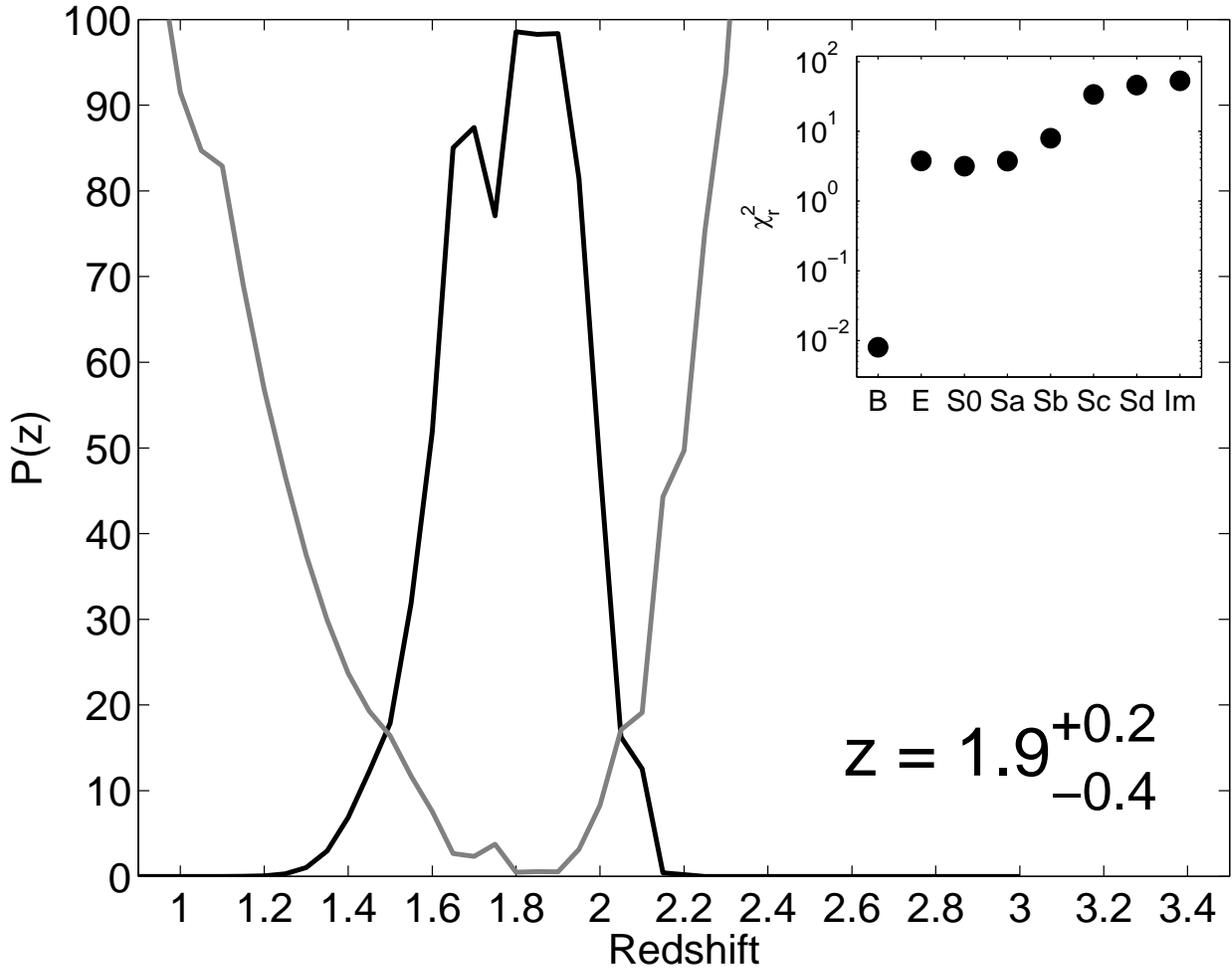,width=6.5in}}
\caption{Probability distribution (black line) and reduced $\chi^2$
(gray line) for the host galaxy photometric redshift calculated with
the {\tt hyperz} software package \citep{bmp00} using a starburst
template.  The best fit redshift is $z=1.9^{+0.2}_{-0.4}$.  The inset
shows the reduced $\chi^2$ for the best fit solution using the various
galaxy templates.  Clearly, the starburst template provides the only
adequate fit to the data.
\label{fig:chi2}}
\end{figure}

The optical/NIR spectral energy distribution of the host galaxy is
shown in Figure~\ref{fig:sed}.  With $R\!-\!K_s=6.2\pm 0.2$ mag the
host is classified as an ERO, and is the reddest GRB host to date.
EROs fall in two general categories of old, passively-evolving
elliptical galaxies, and dust-obscured star forming galaxies
\citep{mcc04}.  Long GRBs have never been localized to elliptical
galaxies, and we further distinguish the two possibilities using the
observed SED.  We use the {\tt hyperz} package \citep{bmp00} to fit a
range of model templates (starburst, E, S0-Sc, and Irr) with the
redshift, stellar population age, and extinction as free parameters.
We assume a \citet{cab+00} extinction curve.  The elliptical galaxy
model provides an unsatisfactory fit with a probability of only
$0.7\%$ ($\chi^2_r=4.1$; Figures~\ref{fig:sed} and \ref{fig:chi2}),
due to the expected flat slope (in $F_\lambda$) beyond 1 $\mu$m and a
brighter $g'$-band magnitude.

The starburst template, on the other hand, provides an excellent fit
($P_{\rm max}=99.9\%$; Figure~\ref{fig:chi2}).  The best-fit
parameters are $z=1.9^{+0.2}_{-0.4}$, an age of about 0.7 Gyr, a
rest-frame extinction, $A_V\approx 0.5$ mag, and an absolute
rest-frame magnitude, $M_{\rm AB}(B)=-23.5\pm 0.1$ mag.  This
corresponds to $L\approx 5$ L$^*$ \citep{dms+05,wfk+06}.  A comparison
to UV-selected galaxies at $z\sim 2$ \citep{sse+05} indicates that the
host mass is $M_*\sim 10^{11-12}$ M$_\odot$.  Using the relation of
\citet{ken98} and the observed $R$-band (2200 \AA\ rest-frame) flux of
$0.46$ $\mu$Jy, we estimate an unobscured star formation rate of $\sim
6$ M$_\odot$ yr$^{-1}$.  The value corrected for extinction is about
an order of magnitude larger.

\begin{figure}
\epsscale{1}
%\plotone{fig4.ps}
\centerline{\psfig{file=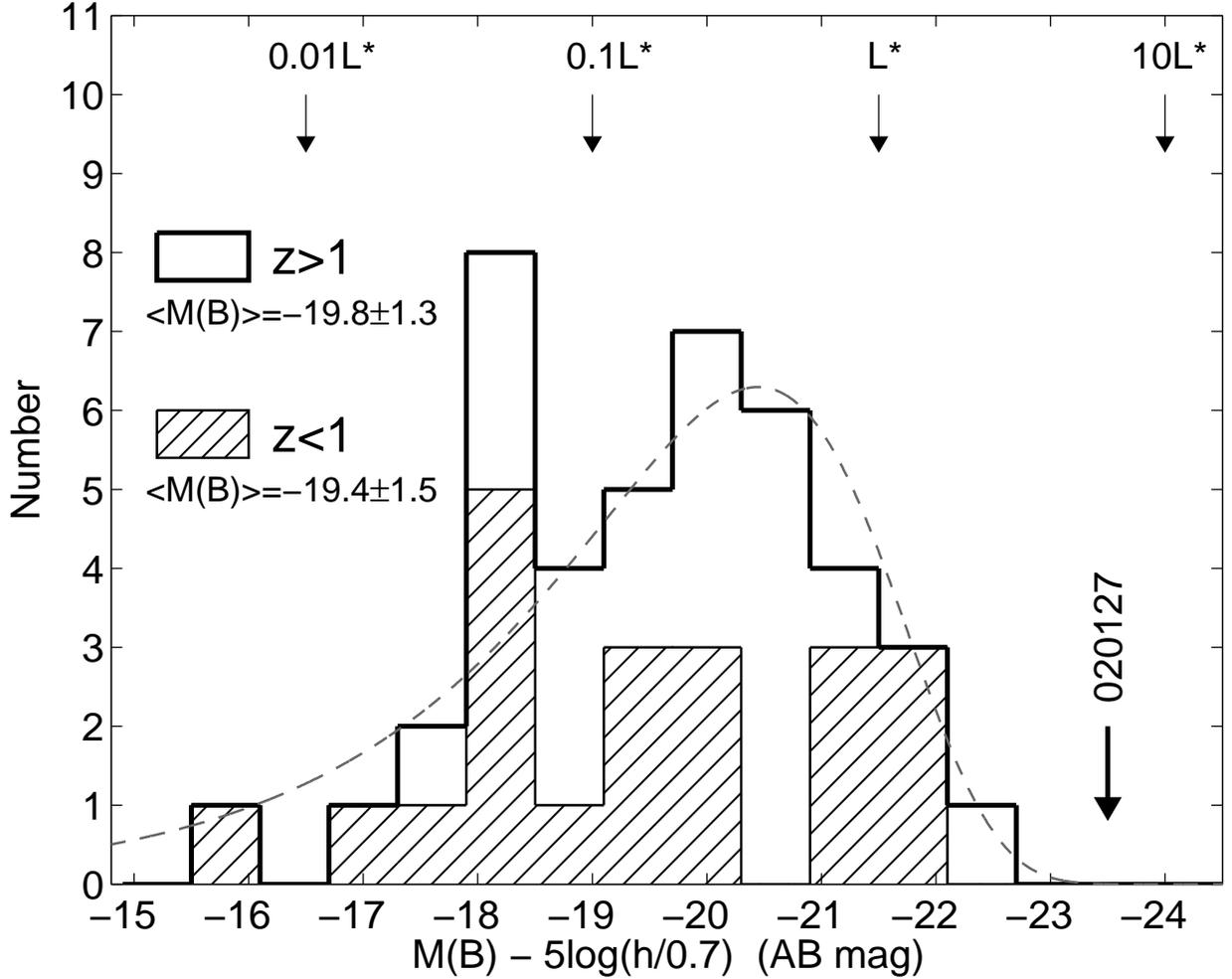,width=6.5in}}
\caption{Distribution of absolute rest-frame $B$-band magnitudes for
GRB host galaxies (Berger et al. in prep.).  The median value for the
sample is $M_{\rm AB}(B)=-19.5$ mag, or $0.1$ L$^*$, and it does not
appear to significantly vary between $z<1$ and $z>1$ (values in the
figure are the median and standard deviation).  \grb\ clearly
stands out as the most luminous, and thus most massive host galaxy
detected to date, but we note that the overall luminosity distribution
is in good agreement with that of other high redshift galaxy sample
(see also \citealt{jbf+05}).  The dashed curve is a representative
Schechter luminosity function with $M_B^*=-21$ mag and $\alpha=-1.4$.
\label{fig:mb}}
\end{figure}

The host of \grb\ is distinguished from other GRB host galaxies in
several ways.  First, it is more than 3 magnitudes redder than the
average value for GRB hosts, and even a magnitude redder than the host
of GRB\,030115 (Figure~\ref{fig:r-k}).  This is due to the combined
effect of extinction and a more evolved stellar population compared to
the typical value of about 0.1 Gyr for other GRB hosts \citep{chg04}.
Second, it is four magnitude more luminous than the median value of
$M_{\rm AB}(B)\approx -19.5$ mag for GRB hosts (with a range of $-16$
to $-22$ mag; Figure~\ref{fig:mb}).  Third, the inferred stellar mass
is nearly two orders of magnitudes larger than the median value of
$10^{9.5}$ M$_\odot$ for GRB hosts \citep{chg04,sgl06}.  Finally, the
unobscured specific star formation rate of about 1 M$_\odot$ yr$^{-1}$
$(L/L*)^{-1}$, is nearly an order of magnitude lower than for other
GRB hosts \citep{chg04}, but we note that with the extinction
correction it is in fact similar.

Perhaps most importantly, a comparison to the mass-metallicity
relation of UV-selected galaxies at $z\sim 2$ \citep{esp+06} indicates
that the host of \grb\ has $12+{\rm log}\,({\rm O/H})\approx 8.6$, or
about $0.5$ Z$_\odot$.  This is also similar to the metallicities of
the so-called distant red galaxies (DRGs; $J-K>2.3$ mag), which have
median ages and masses of $\sim 2$ Gyr and $\sim 10^{11}$ M$_\odot$,
respectively \citep{fvf+04,vff+04}.

\section{Discussion}
\label{sec:disc}

The host galaxy of \grb, identified through X-ray and radio afterglow
observations, is the reddest, and most luminous and massive GRB host
discovered to date.  Unfortunately, the early optical limits
(\S\ref{sec:obs}) are too shallow to ascertain whether the afterglow
was significantly extinguished by dust; from the X-ray flux at 4.14 d
we expect $R\sim 21$ mag at the time of the early optical
observations, at the level of the available limits.  For GRB\,030115,
the only other burst with an ERO host galaxy, the afterglow itself was
shown to be significantly dust extinguished, with $R\!-\!K\approx 6$
mag \citep{lfr+06}.  It is therefore possible that dusty galaxies are
under-represented in the current GRB host sample, due to obscuration 
of the optical afterglow, but we stress that the hosts of several 
dark bursts are not significantly redder than the median of the 
sample \citep{bck+03}.

The discovery of a GRB in a massive, and hence metal-enriched,
starburst galaxy indicates that the proposed low metallicity bias of
GRB progenitors ($\lesssim 0.1$ Z$_\odot$; \citealt{fls+06,sgb+06}) is
not likely to be as severe as previously claimed, particularly at
$z\gtrsim 1$.  This appears to be true not only in terms of an
absolute metallicity cutoff, but also relative to the mass-metallicity
relation of galaxies at high redshift.  In particular, while the hosts
of GRBs at $z\lesssim 0.2$ appear to have metallicities at the low end
of the distribution for local galaxies \citep{sgb+06}, at least some
GRBs at $z\sim 2$ occur in the most metal-enriched galaxies {\it at
that redshift}.  This conclusion is also supported by the
metallicities derived from afterglow absorption spectra (at $z>2$),
which range up to solar values \citep{bpc+06,pro06}, as well as by the
detection of submillimeter emission from some GRB hosts
\citep{bck+03}.  Since GRB progenitors appear to occur at least up to
$\sim 0.5$ Z$_\odot$, and given that the $M$-$Z$ relations at $z\sim
1$ and $\sim 2$ are systematically lower by about 0.3 and 0.6 dex,
respectively, compared to the local relation
\citep{thk+04,sgl+05,esp+06}, we conclude that even if a slight low
metallicity bias does exist, its effect will diminish beyond $z\sim
1$.

\begin{figure}
\epsscale{1}
%\plotone{fig3.ps}
\centerline{\psfig{file=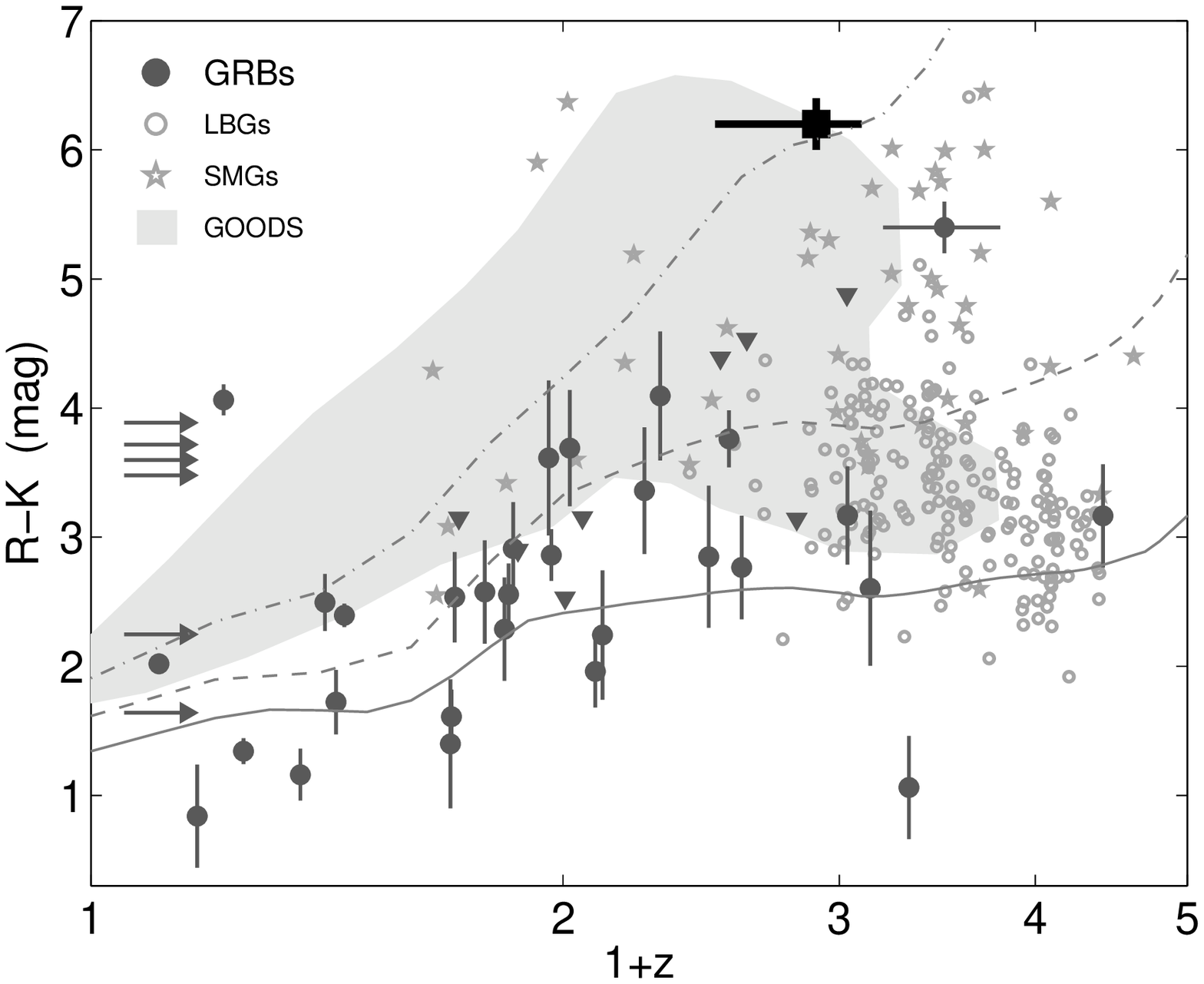,width=6.5in}}
\caption{Apparent $R\!-\!K_s$ color plotted versus redshift for GRB
host galaxies (filled circles; \citealt{cba02,fdm+03,bck+03,lfr+06}),
LBGs (open circles; \citealt{ssa+01,ssp+04}), submillimeter galaxies
(stars; \citealt{scb+04}), and GOODS galaxies (shaded;
\citealt{smm+04}).  Arrows designate GRB hosts without a measured
redshift.  The host galaxy of \grb\ (black square) is significantly 
redder than the typical color of $R\!-\!K_s\approx 2.5-3$ mag.  The 
gray lines are tracks of $R\!-\!K_s$ color as a function of redshift 
from the population synthesis code of \citet{bc03} for a stellar 
population age of 0.1 Gyr (solid), 0.3 Gyr (dashed), and 1 Gyr 
(dot-dashed).  The majority of the GRB hosts appear to have ages of 
$\sim 0.1-0.3$ Gyr.  These young stellar population ages, and not 
metallicity, may be the underlying reason for the blue colors and low
luminosities of GRB hosts.
\label{fig:r-k}}
\end{figure}

Instead, the blue colors of GRB host galaxies may reflect their young
stellar populations.  This is likely related to the fact that GRB
progenitors are massive stars, which explode within a few million
years of formation, thereby leading to the selection of young
starburst galaxies.  In fact, for $z\sim 1-2$, the average $R-K$ color
of a 0.1 Gyr population is about 2.5 mag, in good agreement with the
observed colors of GRB hosts, while for 0.3 and 1 Gyr populations it
is about 1 and 2.5 magnitudes redder, respectively
(Figure~\ref{fig:r-k}).  

This explanation might also account for the transition at low redshift 
($z\lesssim 0.2$) from a representative to a predominantly low 
luminosity host population, since locally young starburst activity 
occurs primarily in low mass galaxies.  \citet{bcw+04} show that 
locally the fraction of galaxies with recent starburst activity 
increases strongly with decreased mass.  Similarly, \citet{bpw+05}
show that while at $z\sim 0.7$ nearly half of all galaxies with 
$M\gtrsim 10^{10}$ M$_\odot$ undergo intense star formation activity,
locally, this number is less than $1\%$.  In this scenario, the 
driving parameter is stellar population age, which may be
misidentified as a low metallicity bias.  Clearly, a study of the
stellar population ages of the low redshift GRB hosts is required to
assess whether age, and not metallicity, in fact plays the underlying
role.

Thus, we conclude that at least at $z\gtrsim 1$, GRBs and their host
galaxies are likely to trace star formation in a relatively unbiased
way, with possibly a preference for younger starburst populations.  At
$z\sim 0$ there seems to be a bias in favor of low luminosity/mass
galaxies \citep{sgb+06}, but this may be driven by stellar population
age and not metallicity.  The luminosity function of GRB hosts (see
also \citealt{jbf+05}) therefore indicates that the bulk of the star
formation at $z\gtrsim 1$ takes place in galaxies fainter than L$^*$,
in good agreement with the steep faint-end slope of the luminosity
function of high redshift galaxies.

\acknowledgments We thank J.~Kollmeier for helpful comments on the
manuscript.  E.B.~is supported by NASA through Hubble Fellowship grant
HST-01171.01 awarded by STSCI, which is operated by AURA, Inc., for
NASA under contract NAS5-26555.  The data presented herein were
obtained at the W.M. Keck Observatory, which is operated as a
scientific partnership among the California Institute of Technology,
the University of California and the National Aeronautics and Space
Administration. The Observatory was made possible by the generous
financial support of the W.M. Keck Foundation.

%\bibliographystyle{apj} 
%\bibliography{journals_apj,refs}

\clearpage

\clearpage

\clearpage

\clearpage

\end{document}